\documentclass[preprint]{aastex}

\usepackage{graphicx}

\renewcommand{\vec}[1]{ {\bf #1} }
\newcommand{\eqref}[1]{(\ref{#1})}

\begin{document}

\title{Spin Tilts in the Double Pulsar Reveal Supernova Spin
  Angular-Momentum Production}

\author{Will M. Farr \and Kyle Kremer}
\affil{Center for Interdisciplinary Exploration and Research in Astrophysics (CIERA)\\
Department of Physics and Astronomy\\
Northwestern University, 2145 Sheridan Road, Evanston, IL 60208}
\email{w-farr@northwestern.edu, kylekremer2012@u.northwestern.edu}

\and

\author{Maxim Lyutikov}
\affil{Physics Department, Purdue University\\
  525 Northwestern Avenue, West Lafayette, IN 47907}
\email{lyutikov@purdue.edu}

\and

\author{Vassiliki Kalogera}
\affil{Center for Interdisciplinary Exploration and Research in Astrophysics (CIERA)\\
Department of Physics and Astronomy\\
Northwestern University, 2145 Sheridan Road, Evanston, IL 60208}
\email{vicky@northwestern.edu}

\begin{abstract}
  The system PSR J0737-3039 is the only binary pulsar known to consist
  of two radio pulsars (PSR J0737-3039 A and PSR J0737-3039 B).  This
  unique configuration allows measurements of spin orientation for
  both pulsars: pulsar A's spin is tilted from the orbital angular
  momentum by \emph{no more than} 14 degrees at 95\% confidence;
  pulsar B's by $130\pm 1$ degrees at $99.7$\% confidence.  This
  spin-spin misalignment requires that the origin of most of B's
  present-day spin is connected to the supernova that formed pulsar B.
  Under the simplified assumption of a single, instantaneous kick
  during the supernova, the spin could be thought of as originating
  from the off-center nature of the kick, causing pulsar B to tumble
  to its misaligned state.  With this assumption, and using current
  constraints on the kick magnitude, we find that pulsar B's
  instantaneous kick must have been displaced from the center of mass
  of the exploding star by at least 1 km and probably 5--10 km.
  Regardless of the details of the kick mechanism and the process that
  produced pulsar B's current spin, the measured spin-spin
  misalignment in the double pulsar system provides 
  an empirical, direct constraint on the angular momentum production in
  this supernova.  This constraint can be used to guide core-collapse
  simulations and the quest for understanding the spins and kicks of
  compact objects.
\end{abstract}

\keywords{pulsars: individual (J0737-3039) --- supernovae: general}

\maketitle

\section{Introduction}

The radio-pulsar system PSR J0737-3039 is the only binary pulsar known
to consist of two radio pulsars: PSR J0737-3039 A \citep{Burgay2003}
and PSR J0737-3039 B \citep{Lyne2004}.  Table
\ref{tab:system-parameters} gives the parameters of this system.  This
unique configuration has permitted measurements of spin orientation
for both pulsars \citep{Ferdman2008,Lyutikov2005,Breton2008}: pulsar
A's spin is tilted from the orbital angular momentum vector by
\emph{no more than} 14 degrees at 95\% confidence \citep{Ferdman2008};
pulsar B's by $130.0^{+1.4}_{-1.2}$ degrees at $99.7$\% confidence.
Here we argue that this large difference between the two pulsar spin
tilts requires that the origin of most of B's spin is connected to its
supernova (SN) explosion; the spin of B's progenitor, expected to be
aligned with the pre-SN orbit due to tidal interactions, cannot be
invoked to explain the present-day misaligned spin of pulsar B.  PSR
J0737-3039 B is currently believed to have formed from an
electron-capture supernova triggered in a massive O-Ne-Mg white dwarf
\citep{vanDenHeuvel2004,Willems2004,Piran2005,Stairs2006,Wang2006,Willems2006,vanDenHeuvel2007,Breton2008,Wong2010}.
Our results demonstrate that, whatever the details of its formation
mechanism, the supernova that formed PSR J0737-3039 B produced the
majority of its current spin.  If the source of the present-day spin
of pulsar B is a single, impulsive kick, then this kick must be
off-center so that it tumbles the pulsar to its current orientation.
Using constraints on the SN kick magnitude derived from the orbital
and kinematic parameters of the system \citep{Wong2010} we find that
this kick must have been displaced from the center of mass of the
exploding star by at least 1 km and probably 5--10 km.  Such offset
distances are a significant fraction of the expected radii of neutron
stars.  Off-center kicks were first suggested in \citet{Spruit1998} on
purely theoretical grounds.

\section{Evolutionary History}

PSR J0737-3039 likely evolved from two stars originally massive enough
to undergo supernova (SN) explosions and form two neutron stars
\citep{Tauris2006} at the end of their nuclear lifetimes.  Given the
measured spin magnitudes and inferred magnetic fields
\citep{Burgay2003,Lyne2004,Ferdman2008,Lyutikov2005,Breton2008},
pulsar A was the first-born neutron star, while pulsar B formed in a
second SN.  After the first SN the system passed through a high-mass
X-ray binary phase.  In this phase, pulsar A accreted matter from its
companion, leading to some spin-up.  Eventually, pulsar A's companion
evolved off the main sequence and its expanding hydrogen envelope
enveloped pulsar A.  In this common-envelope phase, tidal interactions
between the stars circularized the orbit and are expected to have
aligned the spins of pulsar A and of pulsar B's progenitor with the
orbital angular momentum axis (perpendicular to the orbital plane).
The transfer of orbital kinetic energy to the envelope eventually
removed the outer layers of pulsar B's progenitor, leaving pulsar A in
a tight orbit with the exposed helium-rich core of B's progenitor.
After another brief period of mass transfer onto pulsar A
\citep{Dewi2004,Willems2004}, the helium star exploded in the second
SN, forming pulsar B. As a result of the multiple mass-transfer phases
between the two SN events, just before pulsar B's SN the system was in
a close, circular orbit with \emph{both stars' spins aligned with the
  orbital angular momentum vector}.

\begin{figure}
\begin{center}
\plotone{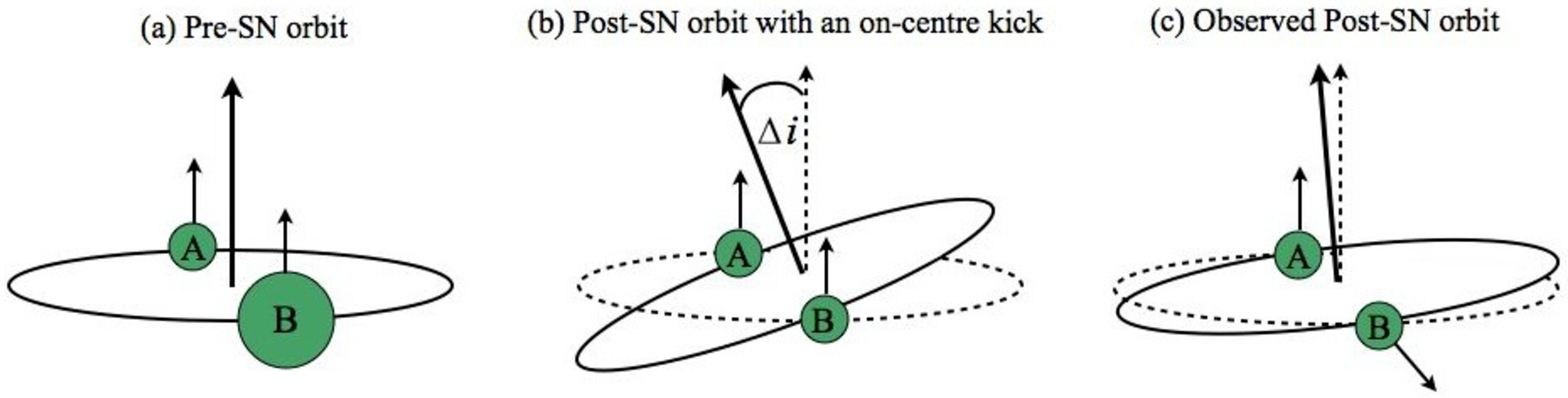}
\end{center}
\caption{\label{fig:system-geometry} Effect of SN kick on binary
  orbit. The pre-SN orbit containing pulsar A and pulsar B's
  progenitor is shown in (a). The effect of an on-center SN kick that
  slightly changes the inclination of the orbit is illustrated in
  (b). Notice the post-SN alignment of the two pulsars' spin
  axes. Part (c) illustrates the present-day orbit with a 130 degree
  misalignment between pulsar B's spin axis and the orbital axis.}
\end{figure}

Due to asymmetries associated with the SN ejecta (matter and/or
neutrinos) SNe are thought capable of imparting a significant recoil
impulse, a ``kick'', to any remnant surviving the explosion (see,
e.g., \citet{Janka2008} and references therein).  When a SN occurs in
a binary system, these kicks can significantly alter the orbital
parameters or even disrupt the binary.  The kick component directed
parallel to the pre-SN orbital plane causes a change in the
eccentricity and semi-major axis of the orbit; the component
perpendicular to the pre-SN orbital plane can also cause a change in
the inclination of the orbital plane.  In the PSR J0737-3039 system,
pulsar A's small spin-tilt angle (less than 14 degrees at 95\%
confidence using a two-pole emission model
\citep{Burgay2003,Lyne2004,Ferdman2008}) is indicative of a relatively
small out-of-plane kick from the SN that formed pulsar B
\citep{Wong2010}.  Pulsar A's spin-orbit misalignment occurs only
because the orbital plane is tilted by the SN kick, while pulsar A's
spin remains fixed in the inertial frame aligned with the pre-SN
orbital angular momentum axis (Figure \ref{fig:system-geometry}).
Such a spin tilt for pulsar A occurs independently of the effects of
the second SN on pulsar B's spin. In other words, the observed tilt of
pulsar A's spin by itself does not require any change in the spin
angular momentum of pulsar B relative to its progenitor.  However,
unless the SN contributes significant amounts of angular momentum to
the nascent pulsar, the orientation of pulsar B's spin will be the
same as its progenitor's spin, i.e.\ aligned with the pre-SN orbital
plane and pulsar A's spin.  Surprisingly, pulsar B's spin is in fact
retrograde: tilted by $130.0^{+1.4}_{-1.2}$ degrees \citep[99.7\%
confidence;][]{Ferdman2008} relative to the current orbital angular
momentum vector (see Figure \ref{fig:system-geometry})!

\section{The Need for Spin Angular Momentum from the Supernova}

To produce pulsar B's retrograde spin, the SN must have significantly
torqued pulsar B, causing it to tumble to the currently observed
spin-orbit orientation.  The pre-SN spin, $\vec{S}_0$, the angular
momentum produced by the SN ejecta, $\Delta \vec{S}$, and the post-SN
spin\footnote{It is important to distinguish between pulsar B's spin
  vector right after the supernova and its present-day spin vector
  because relativistic effects cause the spin vector to precess about
  the total angular momentum \citep{Breton2008}, leading to a
  time-varying azimuthal component.}, $\vec{S}_{SN}$, are related by
the conservation of angular momentum
\begin{equation}
  \label{eq:spins-before-after}
  \vec{S}_{SN} = \vec{S}_0 + \Delta \vec{S}.
\end{equation}

To determine $\Delta \vec{S}$, we must know $\vec{S}_0$ and
$\vec{S}_{SN}$, but we only know the direction, not the magnitude, of
$\vec{S}_0$ and the relationship between $\vec{S}_{SN}$ and the spin
measured today is complicated by relativistic precession
\citep{Breton2008}.  However, we can still place constraints on
$\Delta \vec{S}$.  Relativistic precession causes the individual
pulsar spins to precess about the total angular momentum of the
system, which is approximately parallel to the orbital angular
momentum.  Such precession preserves the angle between the total
angular momentum and the spin (which is the spin colatitude), but not
the azimuthal orientation.  Thus, the colatitude of $\vec{S}_{SN}$
relative to the normal to the current orbital plane is equal to the
colatitude of the current spin---130 degrees.  Based on the spin of
pulsar A, the current orbital plane could be tilted at most $14$
degrees relative to the pre-SN orbital plane.  Therefore the
colatitude of $\vec{S}_{SN}$ relative to the pre-SN orbital
plane---and therefore relative to $\vec{S}_0$---is at least 116
degrees.  This is also the minimum angle between $\vec{S}_0$ and
$\vec{\Delta S}$.  The angular momentum produced by the SN must be
\emph{significantly} mis-aligned with the progenitor spin.  To date,
most SN simulations have focused on non-rotating progenitors
\citep[for example,
see][]{Blondin2007,Rantsiou2011,Wongwathanarat2010}; it remains to be
seen whether the spin produced by the SN from the collapse of a
\emph{rotating} progenitor can be so significantly mis-aligned with
the progenitor's rotation axis.

\begin{figure}
\begin{center}
\plotone{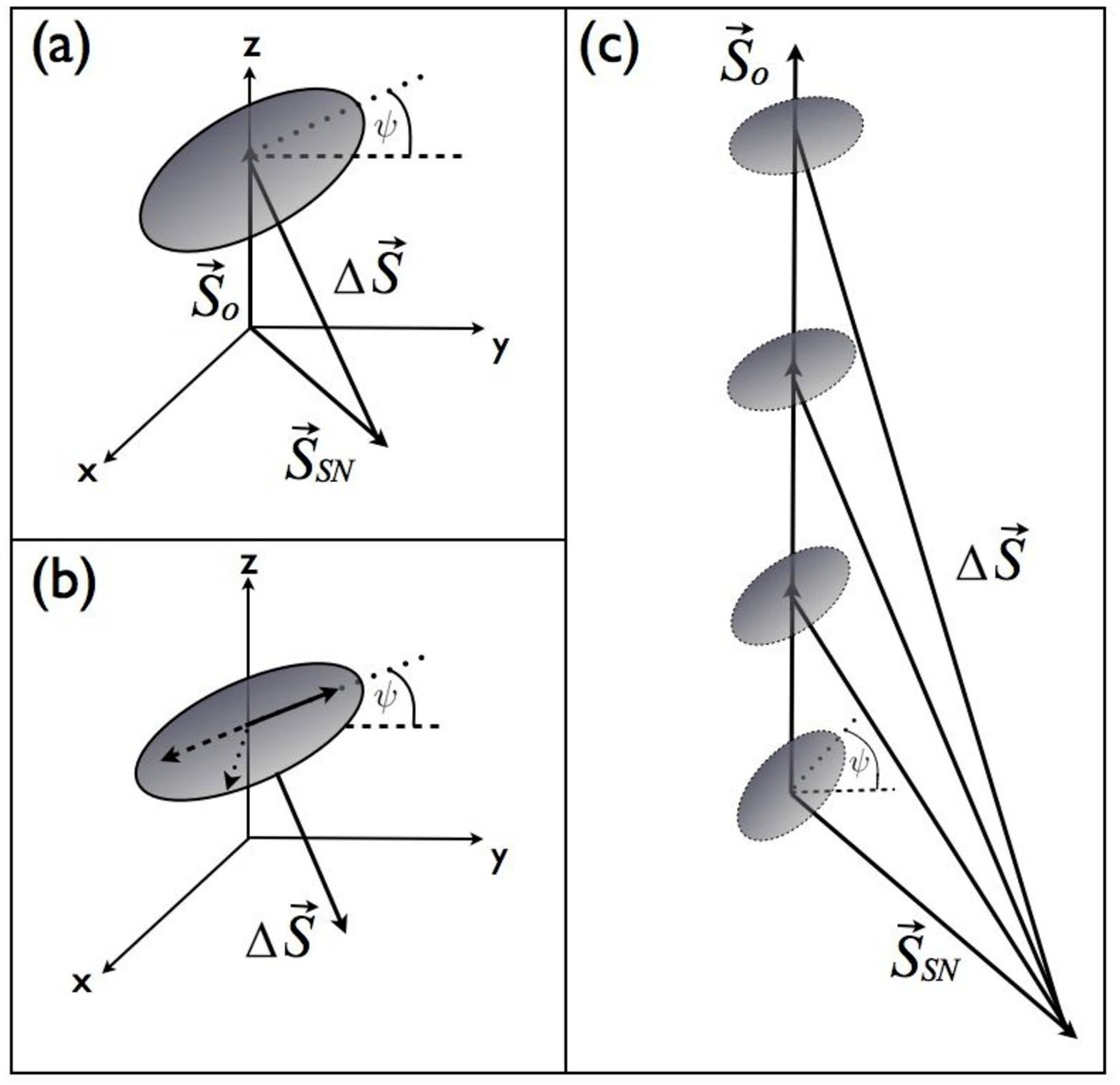}
\end{center}
\caption{\label{fig:spin-geometry} Geometry of the angular momentum
  change due to an off-center kick imparted to the nascent pulsar.
  The $x$-$y$-$z$ reference frame is anchored on pulsar B at the time
  of its SN; the $x$-$y$ plane is the pre-SN orbital plane.  In panel
  (a) we show the relationship between $\vec{S}_0$, $\vec{S}_{SN}$,
  and $\Delta \vec{S}$ and the orientation of the plane orthogonal to
  $\Delta \vec{S}$ (which is inclined by an angle $\psi$ with respect
  to the orbital plane).  In panel (b) we show that vectors lying in
  the plane orthogonal to $\Delta \vec{S}$---like $\vec{v}_K$---can
  have an inclination with respect to the orbital plane that varies
  between $-\psi$ and $\psi$, leading to the constraint on the
  colatitude of $\vec{v}_K$ of $90 - \psi \leq \theta_K \leq 90+ \psi$
  found in Equation (\ref{eq:theta-constraint}).  In panel (c) we show
  that as the magnitude of $\vec{S}_0$ increases, $\Delta \vec{S}$
  increases in magnitude (see Equation \eqref{eq:delta-S-mag}) and
  tilts toward the south pole, reducing $\psi$. Therefore, the most
  conservative constraints on $\theta_K$ are obtained when $S_0 = 0$,
  giving $26 \leq \theta_K \leq 154$ as in Equation
  (\ref{eq:theta-constraint}). }
\end{figure}

The typical moment of inertia \citep{Spruit1998} of a neutron star is
$0.36MR^2$; using pulsar B's measured mass of $1.25 M_{\odot}$ (see
Table \ref{tab:system-parameters}) and a radius of $10$ km, its
current spin angular momentum is $2 \times 10^{45}$ g cm$^2$ s$^{-1}$.
Since this spin is retrograde whereas the pre-SN spin is roughly
aligned (within 14$^\circ$, given pulsar A's small spin tilt) with the
current orbital plane, we can place a lower limit on the change of
angular momentum needed to explain pulsar B's large and retrograde
spin tilt,
\begin{equation}
  \label{eq:delta-S-mag}
  \Delta S \geq 2 \times
  10^{45}\, \textnormal{g} \textnormal{ cm}^2 \textnormal{ s}^{-1},
\end{equation}
where equality holds when $S_0 = 0$.  Because the angle between
$\vec{S}_0$ and $\vec{S}_{SN}$ is greater than 90 degrees, any
progenitor spin only increases the amount of angular momentum that
must be added to the pulsar by the kick.  This is demonstrated
geometrically in Figure \ref{fig:spin-geometry}c.

The above discussion has been fully general.  To extract more
constraints from the observed spin-spin misalignment, we must make
some assumptions about the origin of the pulsar spin.  As a simplified
model to elucidate the scales involved in this scenario, let us assume
that the same impulsive kick (i.e.\ linear momentum) that changes the
orbit of the system is also offset from the center of mass of pulsar
B, and therefore applies a torque sufficient to produce the observed
spin angular momentum.  The kick and offset vectors must lie in the
plane perpendicular to $\Delta \vec{S}$ (see Equation
\ref{eq:spin-change}).  The kick velocity, $\vec{v}_K$, the offset
vector relative to the center of mass, $\vec{r}$, and the change in
B's spin vector are related by
\begin{equation}
  \label{eq:spin-change}
  \Delta \vec{S} = \vec{r} \times \Delta \vec{p} = \vec{r} \times M_B\vec{v}_K,
\end{equation}
where $\Delta \vec{p} = M_B\vec{v}_K$ is the change in linear momentum
induced by a change in velocity of $\vec{v}_K$ in an object with mass
$M_B$.  The offset length $r$ and kick velocity magnitude $v_K$ must
then satisfy the inequality
\begin{equation}
  \label{eq:min-distance}
  r \geq \frac{\Delta S}{M_B v_K},
\end{equation}
where $\Delta \vec{S}$ is the magnitude of the change of B's spin; the
equality holds only when the kick and offset are perpendicular to each
other.

The relative orientation of the current spin provides a constraint on
the kick direction in this scenario.  Let the colatitude of $\Delta
\vec{S}$ relative to the pre-SN orbital plane be $\theta_{\Delta}$;
because the angle between $\vec{S}_0$ and $\vec{S}_{SN}$ is greater
than 90 degrees, no matter the magnitude of $\vec{S}_0$ we must have
$\theta_{\Delta} \geq 116$ degrees.  Let the plane perpendicular to
$\Delta \vec{S}$ make an angle $\psi$ with respect to the pre-SN
orbital plane.  Then we have $\psi = 180 - \theta_{\Delta} \leq 64$
degrees.  The kick, $\vec{v}_K$, lies in this plane and therefore must
have colatitude $\theta_K$ that satisfies
\begin{equation}
\label{eq:theta-constraint}
  90-\psi = 26 \leq \theta_K \leq 154 = 90 +\psi
\end{equation}
This geometry is illustrated in Figure \ref{fig:spin-geometry}.  The
constraint in Equation \ref{eq:theta-constraint} is consistent with
the constraint on kick colatitude in Figure 7 of \citep{Wong2010}, but
tighter.

The constraint on the magnitude of the spin change, Equation
\eqref{eq:delta-S-mag}, together with Equation \eqref{eq:min-distance},
imply a lower limit on the offset distance
\begin{equation}
  \label{eq:min-offset-distance}
  r \geq 3.2 \left( \frac{25 \textnormal{km s}^{-1}}{v_K} \right) \textnormal{km}.
\end{equation}
If we assume that the core of pulsar B's progenitor was in
synchronous, rigid-body rotation just before the supernova then
$S^{SR}_0 \simeq 2 \times 10^{45}\, \textnormal{g}\, \textnormal{cm}^2\,
\textnormal{s}^{-1}$, and the limit on the offset distance rises by a
factor of $1.8$:
\begin{equation}
  r_{SR} \geq 5.8 \left( \frac{25 \textnormal{km s}^{-1}}{v_K} \right) \textnormal{km}.
\end{equation}

\begin{figure}
\begin{center}
\plotone{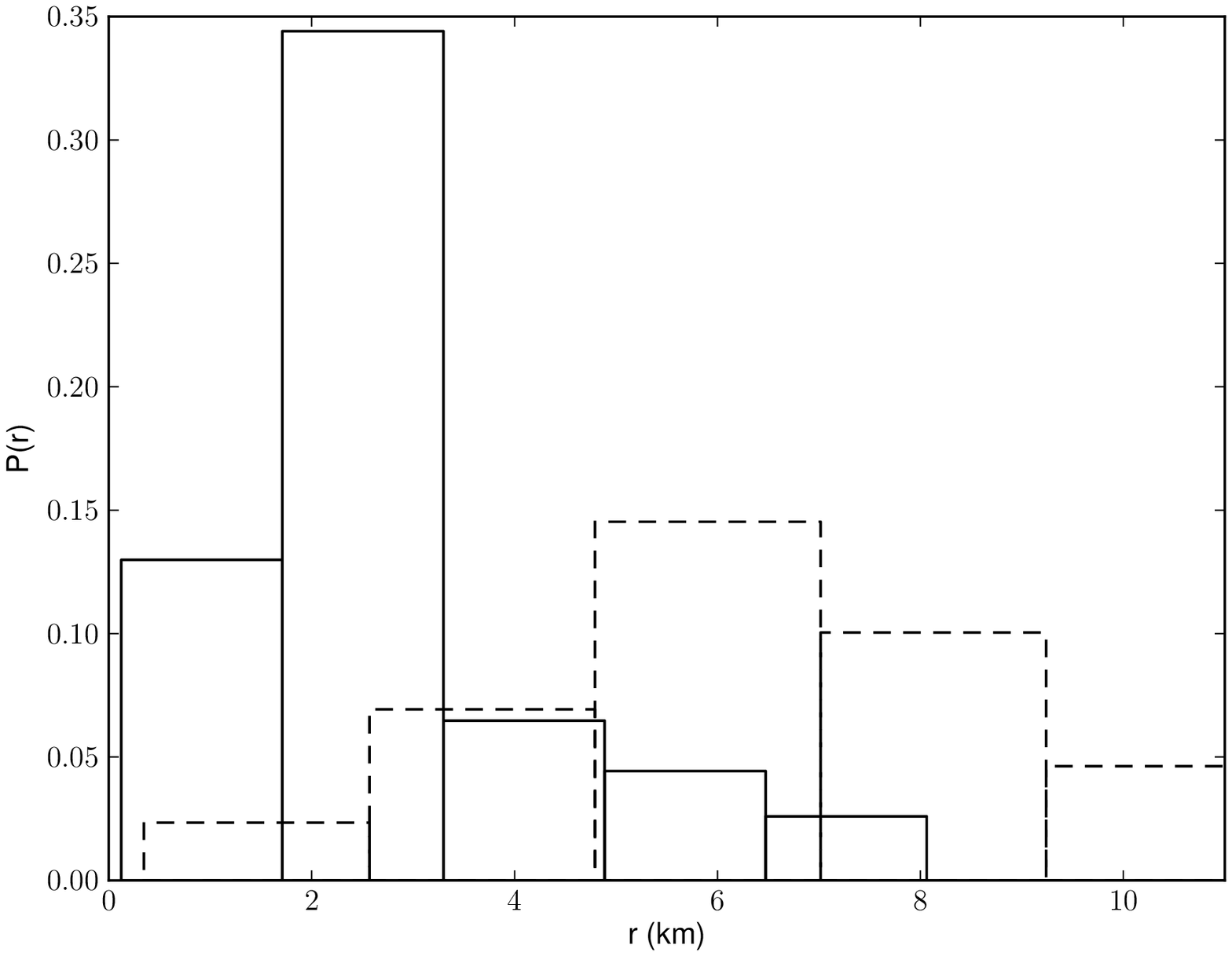}
\end{center}
\caption{\label{fig:offset-pdf} Distributions of minimum offset
  distances implied by the current orbital constraints
  \citep{Wong2010} on kick magnitudes (see Equation
  (\ref{eq:min-offset-distance})).  The solid line gives the
  distribution assuming that $S_0 = 0$, with a kick that is orthogonal
  to the offset vector.  The dashed line assumes that the core of
  pulsar B's progenitor was in synchronous, rigid-body rotation just
  prior to the supernova ($S_0 = S_0^{SR} \simeq 2 \times
  10^{45}\, \textnormal{g}\, \textnormal{cm}^2\, \textnormal{s}^{-1}$)
  and that the kick-offset angle is 40 degrees. }
\end{figure}

\citet{Wong2010} used the measured semi-major axis, eccentricity and
proper motion of the J0737-3039 system (see Table
\ref{tab:system-parameters}) to constrain the kick imparted to the
system by the second SN.  The \citet{Wong2010} analysis assumed that
the pre-SN orbit was circular and that the system came from a
progenitor population with number density
\begin{equation}
  n(R,z) = n_0 \exp \left( -\frac{R}{h_R} \right) \exp \left( -
    \frac{|z|}{h_z} \right),
\end{equation}
with $h_R = 2.8$ kpc and $h_z = 0.07$ kpc the galactic scale length
and height, respectively, moving with the local galactic rotation
velocity.  The SN kick and mass loss must then induce the current
eccentricity and semi-major axis in the orbit and give the system as a
whole a velocity such that it moves in the galactic potential to its
current location in the 100 to 200 Myr since the second SN
\citep{Lorimer2007}.  Because of the uncertainty in the amount of mass
loss, pre-SN semi-major axis, pre-SN galactic location, the age of the
system, and the measurement uncertainty in the current orbital
parameters, a range of kick magnitudes between 0 and 60 km/s is
allowed in the \citet{Wong2010} analysis \citep[Figure 5]{Wong2010}.
In Figure \ref{fig:offset-pdf} we show the probability distribution of
minimum offset distances implied by this distribution of kick
velocities.  Even for large kick velocities, the minimum offset
distance exceeds $\sim 1$ km.  For the smallest allowed kicks, the
minimum offset distances exceed the current $\sim 10$ km radius of the
neutron star.

\begin{table}
  \caption{\label{tab:system-parameters} J07373-3039 System Parameters.  Except as noted, properties are given in \citet{Burgay2003,Lyne2004}. }
\begin{tabular}{|l|l|}
  \hline
  Distance & 600 pc\\
  \hline
  Galactic Latitude & 245.2 deg\\
  \hline 
  Proper Motion & 10 km/s\\
  \hline
  Spin Period (A) & 22.7 ms\\
  \hline
  Spin Period (B) & 2.8 s\\
  \hline
  Mass (A) & 1.34 $M_{\odot}$\\
  \hline
  Mass (B) & 1.25 $M_{\odot}$\\
  \hline
  Spin-orbit misalignment (A) & $\leq$ 14 deg (95\% confidence) \citep{Ferdman2008}\\
  \hline
  Spin-orbit misalignment (B) & $130.0^{+1.4}_{-1.2}$ deg (99.7\%
  confidence) \\
  & \citep{Lyutikov2005,Breton2008}\\
  \hline
  Orbital Period & 2.4 hrs\\
  \hline
  Semi-major Axis & 1.26 $R_{\odot}$\\
  \hline
  Eccentricity & 0.0878\\
  \hline
\end{tabular}
\end{table}

The magnitude of the offset, $r$, required to produce the needed
$\Delta S$ depends on the relative orientation of the offset and kick,
$\vec{v}_K$.  If the angle between $\vec{r}$ and $\vec{v}_K$ in the
plane perpendicular to $\Delta \vec{S}$ is $\theta_{rK}$, then 
\begin{equation}
r = \frac{\Delta S}{M_Bv_K\sin \theta_{rK}},
\end{equation} 
which is larger than the minimum offset distance by a factor of $(\sin
\theta_{rK})^{-1}$ (see Equation \ref{eq:spin-change}). Kicks that are
nearly aligned with the radial vector (small $\theta_{rK}$) require
arbitrarily large off-center distances to match the current spin
orientation.  Even modest misalignments of 40 degrees give an
enhancement factor of $\sin^{-1} 40 \sim 1.6$ over the offset for
perpendicular $\vec{r}$ and $\vec{v}_K$.

\section{Discussion}

Multi-dimensional simulations \citep{Blondin2007} of core-collapse SNe
have shown that an instability in a stationary accretion shock (SASI)
may provide a method of depositing a substantial amount of spin
angular momentum ($2 \times 10^{47}$ g cm$^2$ s$^{-1}$) onto a
proto-neutron star as part of the collapse process itself and separate
from any rotation of the progenitor.  (In fact, the simulations of
\citet{Blondin2007} used non-rotating progenitors.)  This would be
more than enough angular momentum to account for the observed spin
angular momentum of pulsar B ($2 \times 10^{45}$ g cm$^2$ s$^{-1}$).
The general consensus from recent evolutionary studies
\citep{vanDenHeuvel2004,Willems2004,Podsiadlowski2005,Stairs2006,Wang2006,Willems2006,vanDenHeuvel2007,Breton2008,Wong2010}
is that pulsar B was formed in an electron-capture SN, where an iron
core is never formed and instead core collapse is initiated through
electron captures onto Ne/Mg nuclei
\citep{Dessart2006,Kitaura2006,Podsiadlowski2004,Miyaji1980,Nomoto1984,Nomoto1987}.
During the course of such a collapse, the proto-neutron star shrinks
from a radius of $\sim 100$ km to $\sim 10$ km.

We emphasize that the assumption of the foregoing discussion that the
kick is applied at a single location is simplistic (see, e.g.,
\citet{Spruit1998}); in a real supernova, both linear and angular
momentum will be accumulated by the proto-neutron star throughout its
formation at many different locations. In this more realistic context,
the constraints above on offset distances should be interpreted
instead as constraints on the offset scale at which the \emph{bulk} of
the linear and angular momentum is accumulated.  More precise
constraints will require detailed modeling of the hydrodynamic process
of momentum accumulation in the supernova that formed PSR J0737-3039B.
Nevertheless, it is interesting that the location of kicks inferred
from such a simple model is consistent with kick origins in the bulk
of the shrinking proto-neutron star during the supernova (see Figure
\ref{fig:offset-pdf}).  Some recent SN modeling suggests that the
processes that produce the kick and those that impart rotation to the
resulting neutron star produce independent kicks and spins, and therefore there is
little correlation between the kick magnitude and direction and the
rotation imparted to the post-SN compact object
\citep{Wongwathanarat2010,Rantsiou2011}.  In this case the offset
length scale inferred above from the dynamical constraints on the kick
would not be relevant.  We conclude that \emph{only if} pulsar B's
spin is actually linked to the torque induced by the physical
mechanism producing the kick it must be offset from the center of mass
of the collapsing neutron star progenitor.

Regardless of the specifics of the collapse process, however, the
expected alignment of the spin of pulsar B's SN progenitor with the
pre-SN orbital angular momentum and the observed misalignment of
pulsar B's spin and orbit at present uniquely imply that pulsar B's
spin is dominated by angular momentum produced during the SN process,
not angular momentum provided by the progenitor.  The realization of
this empirical constraint on angular momentum production in supernovae
presented here is uniquely enabled by the spin spin misalignment in
the PSR J0737-3039 system \citep{Lyne2004,Ferdman2008,Lyutikov2005}
and can be used to guide core-collapse simulations and the quest for
the understanding of compact object spins and kicks.

\acknowledgements 

We thank Tsing Wai Wong for providing data on the distribution of
allowed pulsar kicks from the current orbital constraints from
\citet{Wong2010} (Figure 5).  We also thank Ingrid Stairs, Hans-Thomas
Janka, Adam Burrows, and Rodrigo Fernandez for helpful comments on
drafts of this manuscript.  WMF and VK are partially supported by NSF
grant AST-0908930.  KK acknowledges support from a NASA summer
research grant through the Illinois Space Grant NNG05G381H.  ML is
supported by NASA grant NNX09AH37G.


\begin{thebibliography}{0}
\expandafter\ifx\csname natexlab\endcsname\relax\def\natexlab#1{#1}\fi

\end{thebibliography}


\begin{thebibliography}{28}
\expandafter\ifx\csname natexlab\endcsname\relax\def\natexlab#1{#1}\fi

\bibitem[{{Blondin} \& {Mezzacappa}(2007)}]{Blondin2007}
{Blondin}, J.~M., \& {Mezzacappa}, A. 2007, \nat, 445, 58,
  arXiv:astro-ph/0611680

\bibitem[{{Breton} {et~al.}(2008){Breton}, {Kaspi}, {Kramer}, {McLaughlin},
  {Lyutikov}, {Ransom}, {Stairs}, {Ferdman}, {Camilo}, \&
  {Possenti}}]{Breton2008}
{Breton}, R.~P. {et~al.} 2008, Science, 321, 104, arXiv:0807.2644

\bibitem[{Burgay {et~al.}(2003)Burgay, D'Amico, Possenti, Manchester, Lyne,
  Joshi, McLaughlin, Kramer, Sarkissian, Camilo, Kalogera, Kim, \&
  Lorimer}]{Burgay2003}
Burgay, M. {et~al.} 2003, \nat, 426, 531

\bibitem[{{Dessart} {et~al.}(2006){Dessart}, {Burrows}, {Ott}, {Livne}, {Yoon},
  \& {Langer}}]{Dessart2006}
{Dessart}, L., {Burrows}, A., {Ott}, C.~D., {Livne}, E., {Yoon}, S., \&
  {Langer}, N. 2006, \apj, 644, 1063, arXiv:astro-ph/0601603

\bibitem[{Dewi \& van~den Heuvel(2004)}]{Dewi2004}
Dewi, J. D.~M., \& van~den Heuvel, E. P.~J. 2004, \mnras, 349, 169,
  arXiv:astro-ph/0312152

\bibitem[{{Ferdman} {et~al.}(2008){Ferdman}, {Stairs}, {Kramer}, {Manchester},
  {Lyne}, {Breton}, {McLaughlin}, {Possenti}, \& {Burgay}}]{Ferdman2008}
{Ferdman}, R.~D. {et~al.} 2008, in American Institute of Physics Conference
  Series, Vol. 983, 40 Years of Pulsars: Millisecond Pulsars, Magnetars and
  More, ed. {C.~Bassa, Z.~Wang, A.~Cumming, \& V.~M.~Kaspi}, 474--478,
  arXiv:0711.4927

\bibitem[{{Janka} {et~al.}(2008){Janka}, {Marek}, {M{\"u}ller}, \&
  {Scheck}}]{Janka2008}
{Janka}, H., {Marek}, A., {M{\"u}ller}, B., \& {Scheck}, L. 2008, in American
  Institute of Physics Conference Series, Vol. 983, 40 Years of Pulsars:
  Millisecond Pulsars, Magnetars and More, ed. {C.~Bassa, Z.~Wang, A.~Cumming,
  \& V.~M.~Kaspi}, 369--378, arXiv:0712.3070

\bibitem[{{Kitaura} {et~al.}(2006){Kitaura}, {Janka}, \&
  {Hillebrandt}}]{Kitaura2006}
{Kitaura}, F.~S., {Janka}, H., \& {Hillebrandt}, W. 2006, \aap, 450, 345,
  arXiv:astro-ph/0512065

\bibitem[{{Lorimer} {et~al.}(2007){Lorimer}, {Freire}, {Stairs}, {Kramer},
  {McLaughlin}, {Burgay}, {Thorsett}, {Dewey}, {Lyne}, {Manchester}, {D'Amico},
  {Possenti}, \& {Joshi}}]{Lorimer2007}
{Lorimer}, D.~R. {et~al.} 2007, \mnras, 379, 1217, arXiv:0705.3269

\bibitem[{Lyne {et~al.}(2004)Lyne, Burgay, Kramer, Possenti, Manchester,
  Camilo, McLaughlin, Lorimer, D'Amico, Joshi, Reynolds, \& Freire}]{Lyne2004}
Lyne, A.~G. {et~al.} 2004, Science, 303, 1153

\bibitem[{{Lyutikov} \& {Thompson}(2005)}]{Lyutikov2005}
{Lyutikov}, M., \& {Thompson}, C. 2005, \apj, 634, 1223, arXiv:astro-ph/0502333

\bibitem[{{Miyaji} {et~al.}(1980){Miyaji}, {Nomoto}, {Yokoi}, \&
  {Sugimoto}}]{Miyaji1980}
{Miyaji}, S., {Nomoto}, K., {Yokoi}, K., \& {Sugimoto}, D. 1980, \pasj, 32, 303

\bibitem[{{Nomoto}(1984)}]{Nomoto1984}
{Nomoto}, K. 1984, \apj, 277, 791

\bibitem[{{Nomoto}(1987)}]{Nomoto1987}
------. 1987, \apj, 322, 206

\bibitem[{{Piran} \& {Shaviv}(2005)}]{Piran2005}
{Piran}, T., \& {Shaviv}, N.~J. 2005, \prl, 94, 051102, arXiv:astro-ph/0409651

\bibitem[{{Podsiadlowski} {et~al.}(2005){Podsiadlowski}, {Dewi}, {Lesaffre},
  {Miller}, {Newton}, \& {Stone}}]{Podsiadlowski2005}
{Podsiadlowski}, P., {Dewi}, J.~D.~M., {Lesaffre}, P., {Miller}, J.~C.,
  {Newton}, W.~G., \& {Stone}, J.~R. 2005, \mnras, 361, 1243,
  arXiv:astro-ph/0506566

\bibitem[{{Podsiadlowski} {et~al.}(2004){Podsiadlowski}, {Langer},
  {Poelarends}, {Rappaport}, {Heger}, \& {Pfahl}}]{Podsiadlowski2004}
{Podsiadlowski}, P., {Langer}, N., {Poelarends}, A.~J.~T., {Rappaport}, S.,
  {Heger}, A., \& {Pfahl}, E. 2004, \apj, 612, 1044, arXiv:astro-ph/0309588

\bibitem[{Rantsiou {et~al.}(2011)Rantsiou, Burrows, Nordhaus, \&
  Almgren}]{Rantsiou2011}
Rantsiou, E., Burrows, A., Nordhaus, J., \& Almgren, A. 2011, \apj, 732, 57,
  arXiv:1010.5238

\bibitem[{{Spruit} \& {Phinney}(1998)}]{Spruit1998}
{Spruit}, H., \& {Phinney}, E.~S. 1998, \nat, 393, 139, arXiv:astro-ph/9803201

\bibitem[{{Stairs} {et~al.}(2006){Stairs}, {Thorsett}, {Dewey}, {Kramer}, \&
  {McPhee}}]{Stairs2006}
{Stairs}, I.~H., {Thorsett}, S.~E., {Dewey}, R.~J., {Kramer}, M., \& {McPhee},
  C.~A. 2006, \mnras, 373, L50, arXiv:astro-ph/0609416

\bibitem[{{Tauris} \& {van den Heuvel}(2006)}]{Tauris2006}
{Tauris}, T.~M., \& {van den Heuvel}, E.~P.~J. 2006, {Formation and evolution
  of compact stellar X-ray sources}, ed. {Lewin, W.~H.~G.~\& van der Klis, M.},
  623--665

\bibitem[{{van den Heuvel}(2004)}]{vanDenHeuvel2004}
{van den Heuvel}, E.~P.~J. 2004, in ESA Special Publication, Vol. 552, 5th
  INTEGRAL Workshop on the INTEGRAL Universe, ed. {V.~Schoenfelder, G.~Lichti,
  \& C.~Winkler}, 185--+, arXiv:astro-ph/0407451

\bibitem[{{van den Heuvel}(2007)}]{vanDenHeuvel2007}
{van den Heuvel}, E.~P.~J. 2007, in American Institute of Physics Conference
  Series, Vol. 924, The Multicolored Landscape of Compact Objects and Their
  Explosive Origins, ed. {T.~di Salvo, G.~L.~Israel, L.~Piersant, L.~Burderi,
  G.~Matt, A.~Tornambe, \& M.~T.~Menna}, 598--606, arXiv:0704.1215

\bibitem[{{Wang} {et~al.}(2006){Wang}, {Lai}, \& {Han}}]{Wang2006}
{Wang}, C., {Lai}, D., \& {Han}, J.~L. 2006, \apj, 639, 1007,
  arXiv:astro-ph/0509484

\bibitem[{Willems \& Kalogera(2004)}]{Willems2004}
Willems, B., \& Kalogera, V. 2004, \apjl, 603, L101, arXiv:astro-ph/0312426

\bibitem[{{Willems} {et~al.}(2006){Willems}, {Kaplan}, {Fragos}, {Kalogera}, \&
  {Belczynski}}]{Willems2006}
{Willems}, B., {Kaplan}, J., {Fragos}, T., {Kalogera}, V., \& {Belczynski}, K.
  2006, \prd, 74, 043003, arXiv:astro-ph/0602024

\bibitem[{{Wong} {et~al.}(2010){Wong}, {Willems}, \& {Kalogera}}]{Wong2010}
{Wong}, T., {Willems}, B., \& {Kalogera}, V. 2010, \apj, 721, 1689,
  arXiv:1008.2397

\bibitem[{{Wongwathanarat} {et~al.}(2010){Wongwathanarat}, {Janka}, \&
  {M{\"u}ller}}]{Wongwathanarat2010}
{Wongwathanarat}, A., {Janka}, H., \& {M{\"u}ller}, E. 2010, \apjl, 725, L106,
  arXiv:1010.0167

\end{thebibliography}

\end{document}